\documentclass[runningheads]{llncs}
\usepackage[T1]{fontenc}
\usepackage{cite}
\usepackage{graphicx}
\usepackage{booktabs}
\usepackage{multirow}
\usepackage{textcomp}
\usepackage{tabularx}
\begin{document}
\title{A Study on the Importance of Features in Detecting Advanced Persistent Threats Using Machine Learning\thanks{Submitted to CSCI--RTCW 2024.}}
\titlerunning{A Study on the Importance of Features in Detecting APTs}
%
\author{Ehsan Hallaji\inst{1}\orcidID{0000-0002-9956-4003} \and
Roozbeh Razavi-Far\inst{1,2}\orcidID{0000-0002-4330-3656} \and
Mehrdad Saif\inst{1}\orcidID{0000-0002-7587-4189}}

\authorrunning{E. Hallaji et al.}
%
\institute{University of Windsor, Windsor, ON N9B 3P4, Canada \\
\email{\{hallaji, msaif\}@uwindsor.ca}
\and University of New Brunswick, Fredericton, NB E3B 5A3, Canada \\
\email{roozbeh.razavi-far@unb.ca}}
\maketitle              
\begin{abstract}
Advanced Persistent Threats (APTs) pose a significant security risk to organizations and industries. These attacks often lead to severe data breaches and compromise the system for a long time. Mitigating these sophisticated attacks is highly challenging due to the stealthy and persistent nature of APTs. Machine learning models are often employed to tackle this challenge by bringing automation and scalability to APT detection. Nevertheless, these intelligent methods are data-driven, and thus, highly affected by the quality and relevance of input data. This paper aims to analyze measurements considered when recording network traffic and conclude which features contribute more to detecting APT samples. To do this, we study the features associated with various APT cases and determine their importance using a machine learning framework. To ensure the generalization of our findings, several feature selection techniques are employed and paired with different classifiers to evaluate their effectiveness. Our findings provide insights into how APT detection can be enhanced in real-world scenarios.

\end{abstract}

\section{Introduction}
Cyber threats have dramatically become more widespread with the world's ongoing movement toward complete digitization \cite{6198335, 10163904}. The financial loss, reputational damage, and at times compromises in national security caused by cyber-attacks make cybersecurity a growing concern for businesses across all sectors \cite{10.1007/978-3-030-61638-0_16,case2016analysis}. Confidence in digital systems is maintained by protecting digital assets and ensuring the availability of data, integrity, and privacy. Conventional defenses struggle with coping with the rising intensity and scale of cyber-attacks, both in number and sophistication \cite{LI20218176, HALLAJI2024103730,9563211}. Thus, sophisticated and adaptive solutions should be developed that facilitate eliminating complex and large-scale cyber-attacks.

APTs are among the most sophisticated and complex cyber-attacks, making them very challenging to detect \cite{8606252, 10.1145/3386581}. These stealthy and targeted attacks are commonly carried out in multiple steps to infiltrate a system and can remain active for an extended period of time. APT can have a variety of objectives, including stealing sensitive data and disrupting critical infrastructure. These enemies often remain unnoticed as they take advantage of the network and reveal the weaknesses in the system \cite{9923774}. APTs are especially dangerous for both the public and private sectors since they are typically launched by highly skilled and resourceful hacker groups, such as nation-states or organized crime groups.

Many research efforts are dedicated to addressing APTs in computer networks \cite{8941015, HUANG2020101660, 8417919, Hallaji2024}. The majority of these studies are based on intrusion detection systems, network monitoring tools, and firewalls, and they provide baseline security against such adversaries. Recently, machine learning-based systems have been employed to facilitate the detection process by focusing on patterns and anomalies in network traffic, system logs, and user behavior \cite{9947295, 10.1007/978-3-319-60080-2_21, 9695986}. These models provide a dynamic and extensible approach that will enable organizations to automatically detect complex threats with less dependence on signature-based defenses.

Despite the potency of machine learning-based detection systems, the data-driven backbone of these models is highly affected by the quality of data \cite{HASSANI2021104150, 9207066, 9907008}. For instance, redundant or irrelevant features that may be useful for other tasks may increase false alarms and missed detection \cite{
9609642}. In other words, the quality of features in the input space of the detection model directly impacts the detection performance of APT samples \cite{10.1007/978-3-319-60080-2_21}. Moreover, if the employed features do not reflect the system dynamics when it is exposed to APT samples, the machine learning model cannot be effective, regardless of the model's robustness.

To tackle the aforementioned issues, this paper studies the network measurements used in detecting APTs and determines the features contributing the most to this process. To do so, we design a machine learning framework that ensures the generalization of our research findings. This has been ensured by employing several Feature Selection (FS) techniques from different categories and combining them with multiple classifiers. In addition, we study 15 different cases of APTs based on public records to cover a wide range of APT attacks. Our findings will provide insight into optimizing detecting APTs using machine learning techniques, addressing the critical challenge of data quality in APT defense strategies.

The remainder of this paper is organized as follows. Section \ref{sec:background} explains the background and preliminaries. Section \ref{sec:design} explains the methodology of this paper. Experimental results are reported and analyzed in Section \ref{sec:results}. Finally, the paper is concluded in Section \ref{sec:conclusion}.

\section{Background}
\label{sec:background}
Here, we first provide a more detailed explanation of APT. Then, we review popular real-world APT cases considered in this work.

\subsection{Advanced Persistent Threats}
APTs are targeted, intelligence-driven, and complex cyber-attacks launched against high-value targets such as critical infrastructures and financial institutions \cite{8606252}. APTs are multi-step attacks, usually starting with intelligence collection through reconnaissance. Based on the security holes identified during reconnaissance, APT gains access to the system and exploits the system. To ensure continuous access to the targeted system, APTs often inject backdoors into the system so that they can revive the connection when desired. From there, attackers explore the network to extend their access to more sensitive components of the system. Finally, the attacker reaches a malicious objective such as hijacking sensitive data or disrupting critical infrastructures.

\subsection{Cases of Advanced Persistent Threat}
Several APT cases are considered in this study. The name of these cases indicates the hacker group responsible for them. These cases are briefly introduced in the following.

\subsubsection{APT28 (Fancy Bear)} A Russian cyber-espionage group linked to military intelligence. It targets government, military, and media sectors, mainly in Europe and the U.S. It is known for interfering in the 2016 election in the U.S. using spear phishing and sophisticated malware to carry out espionage.

\subsubsection{APT29 (Cozy Bear)} Another highly recognized state-sponsored Russian group linked to the Foreign Intelligence Service. APT29 focuses on long-term espionage campaigns targeting government, diplomatic, and healthcare organizations. It typically performs extremely stealthy and sophisticated malware attacks, like in the case of SolarWinds back in 2020.

\subsubsection{APT30} A sophisticated cyber-espionage group, reportedly supported by China and with a large focus on targets such as Southeast Asian governments, aerospace, and defense industries. Spear phishing and custom malware are just some of the methods that have been utilized since at least 2005 to steal sensitive data.

\subsubsection{Carbanak} A cybercrime gang launched attacks against financial institutions worldwide starting in 2013. It's known to infiltrate bank networks via phishing emails and then use malware to siphon off millions more by manipulating internal systems.

\subsubsection{Desert Falcon} An Arabic-speaking hacker group that has been active since 2011, targets military, media, and political targets for espionage in the Middle East. It conducts sensitive data theft from governments and corporations using malware.

\subsubsection{Hurricane Panda} A Chinese hacker group that mainly targets aerospace and energy. It is well-known for its advanced toolset, featuring zero-day exploits. Hurricane Panda therefore steals proprietary information through cyber espionage.

\subsubsection{Lazarus Group} Some would consider it sponsored by North Korea; it has been accused of several high-profile hacks, including Sony Pictures and the WannaCry ransomware. It performs financial crime and espionage, quite frequently against global financial institutions.

\subsubsection{Mirage} A cyber-espionage group targeting energy companies and government organizations in East Asia. The nature of its hacks is described as phishing campaigns and custom malware aimed at intelligence gathering on behalf of the victims.

\subsubsection{Patchwork} An Indian hacker group targets think tanks, government institutions, and the defense sector throughout South Asia. The malware is often just a rehash of code stolen from other APT groups, and intellectual property theft is their ultimate focus.

\subsubsection{Sandworm} One of the most feared Russian hacker groups because it boasts a large number of destructive attacks. Among them is the highly publicized attack on the 2015 Ukraine power grid. Sandworm deploys highly sophisticated malware and takes a huge interest in targeting vital infrastructure for political reasons.

\subsubsection{Shiqiang} This hacker group appears to be related to Chinese state actors and focuses its operations on cyber-espionage activities targeting the government and industrial sectors around the world. The prime intention behind such threats is to steal intelligence or trade secrets from strategic industries.

\subsubsection{Transparent Tribe} Originating from Pakistan, Transparent Tribe targets Indian government and military verticals. It performs cyber espionage by infecting sensitive systems with malware such as Crimson RAT.

\subsubsection{Violin Panda} An advanced cyber-espionage entity based in China targets organizations in technology and defense industries around the world. The main modus operandi of Violin Panda is spear-phishing campaigns, while custom malware is used for data exfiltration.

\subsubsection{Winnti Group} This Chinese state-sponsored group, also known as Winnti, concentrates on stealing intellectual property through espionage. The primary targets of the group are the gaming and software industries. It is also known to be behind supply chain attacks and to maintain long-term persistence in networks.

\subsection{Case Study}
To simulate real-world APTs in computer networks, we consider the data introduced in \cite{10.1007/978-3-319-60080-2_21}. This data is created based on known APTs with publicly available reports. These data contain 2086 APT samples obtained from 15 different APTs, as listed in Table \ref{tab:attacks}. A total of 9021 non-APT malware are also included in the data. The objective of this case study is to distinguish between APT and non-APT attacks so that malware triage can be performed accordingly. Table \ref{tab:attacks} also indicates the population of samples associated with each APT mechanism. To minimize the latency, samples are recorded using static features that can be extracted with minimal delay. Initially, 4000 features were extracted using the PEFrame tool. These features consist of optional headers, MS-DOS headers, file headers, obfuscated string statistics, Mutex, Packer, buckets, and imported API. 

\begin{table}[h]
\centering
    \caption{APT mechanisms included in the data \cite{10.1007/978-3-319-60080-2_21}. Class populations are indicated accordingly.}
    \setlength{\tabcolsep}{15pt}
    \begin{tabular}{clc}
    \toprule
        Label & Class & \#. Samples \\
        \midrule
        1 & APT28 & 68 \\
        2 & APT29 & 205 \\
        3 & APT30 & 101 \\
        4 & Carbanak & 105\\
        5 & Desert Falcon & 45 \\
        6 & Hurricane Panda & 315\\
        7 & Lazarus Group & 58\\
        8 & Mirage & 54\\
        9 & Patchwork & 559\\
        10 & Sandwork & 44\\
        11 & Shiqiang & 31\\
        12 & Transparent Tribe & 267\\
        13 & Violin Panda & 23\\
        14 & Volatile Cedar & 35\\
        15 & Winnti Group & 176 \\
        16 & Non-APT Malware & 9021\\
        \bottomrule
    \end{tabular}
    \label{tab:attacks}
\end{table}


\section{Methodology}
\label{sec:design}

\begin{figure}[t]
    \centering
    \includegraphics[trim={0.5cm 0.5cm 0.5cm 0.5cm},clip,width=\textwidth]{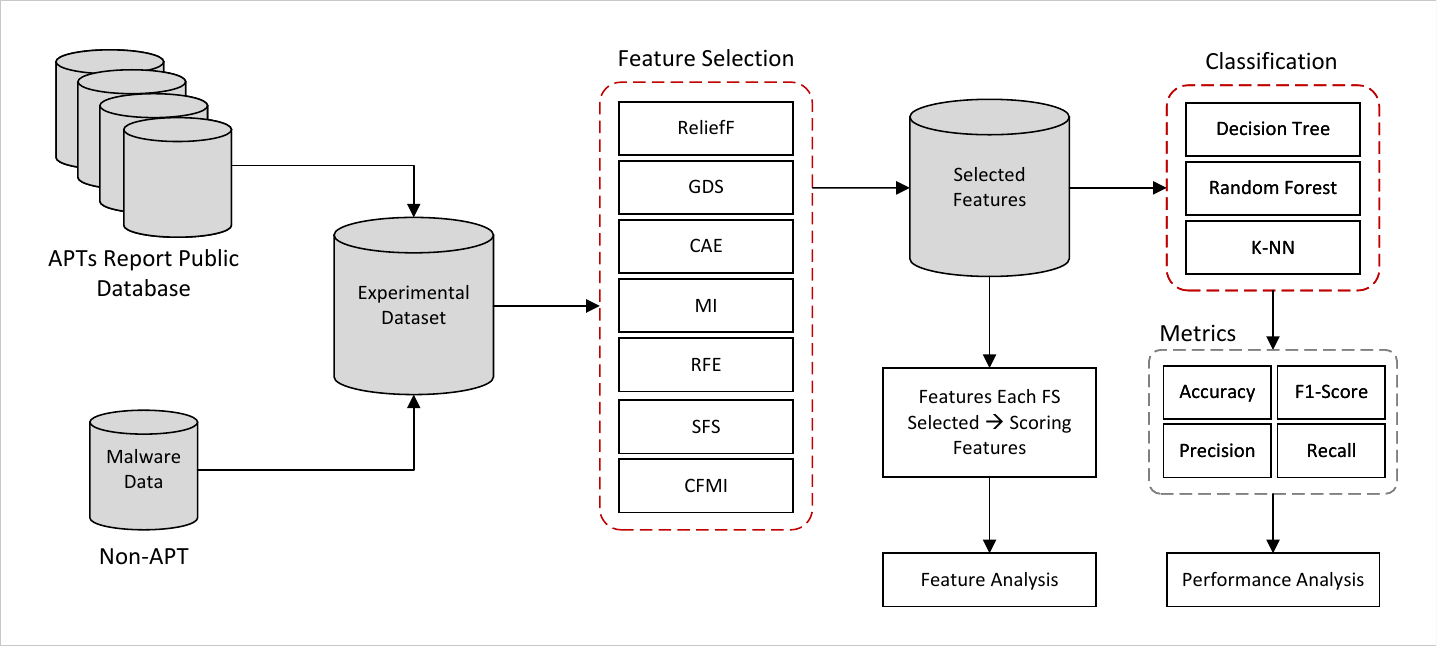}
    \caption{Block diagram of the design feature analysis framework for data-driven detection of malicious samples associated with APT.}
    \label{fig:diagram}
\end{figure}

Fig. \ref{fig:diagram} illustrates the block diagram of the design feature analysis framework for data-driven detection of APT samples. Initially, a dataset is created by combining data associated with different APT cases based on the available public records \cite{10.1007/978-3-319-60080-2_21}. Then, the unified dataset is processed by several FS algorithms that are selected from three main categories, namely filter, wrapper, and embedded FS. The filter approach often uses computationally efficient techniques to score features using statistical measures. ReliefF \cite{relief} and Spectral Feature Selection (SFS) \cite{10.1145/1273496.1273641} both follow this approach and are used within the designed framework.  The wrapper approach uses a machine learning model to iteratively eliminate features based on the prediction performance associated with using them. Recursive Feature Elimination (RFE), Greedy Dynamic Selection (GDS) \cite{pmlr-v202-covert23a}, and FS using Mutual Information (MI) \cite{1453511} are examples of this category that are selected in our framework \cite{rfe}. Embedded FS selects top features during learning, which is advantageous in terms of computational efficiency. Additionally, the classification performance using this approach often outperforms that of the others. From this category, Concrete Autoencoder (CAE) \cite{pmlr-v97-balin19a} and Concrete FS using MI (CFMI) \cite{HASSANI2021104150} are included in the designed analytical framework. 

Each of the FS algorithms processes the prepared dataset and selects 110 features from the total pool of features. These features are then used for two separate studies. Firstly, we track the selected features by each FS method and assign a score to each feature. This score is calculated based on the number of FS techniques that selected a certain feature. Then, features are categorized into different priority groups based on the estimated scores. The priority groups and their features are analyzed and studied afterward.

\begin{figure}[t]
    \centering
    \includegraphics[trim={3.5cm 1cm 4.5cm 1.5cm },clip,width=\textwidth]{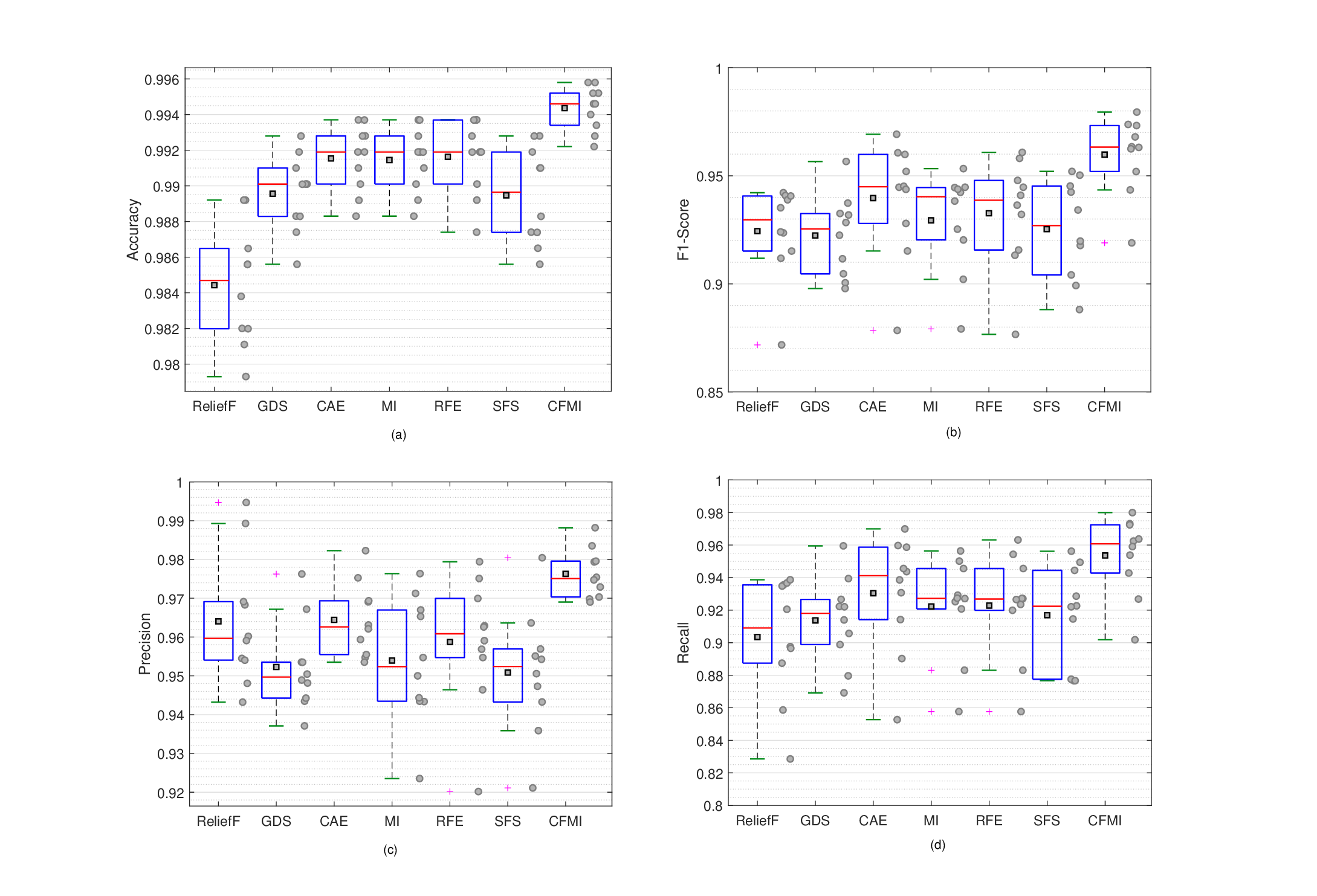}
    \caption{Post-FS performance of APT detection in terms of accuracy, F1-score, precision, and recall. Solid circles indicate obtained measurements in each iteration of cross-validation.}
    \label{fig:box}
\end{figure}

In the second study, the selected features are passed to three classifiers, namely decision tree, random forest, and k-Nearest Neighbors (k-NN). The outputs of these classifiers are averaged to make the results independent of a certain classifier. In this process, four different performance metrics are used to enable studying various aspects of APT detection such as overall accuracy, performance under class imbalance, false alarm rate, and false positive rate. These results are then used to study the potency of FS in enhancing the performance of APT detection.

\section{Experimental Results}
\label{sec:results}
This section initially explains the experimental setting used in the simulations and then proceeds to report and analyze the obtained results. The importance of features for APT detection is studied afterward.

\subsection{Experimental Setting}
Experiments are evaluated in terms of accuracy, F1-score, precision, and recall. To ensure the statistical reliability of the results, experiments are cross-validated using 10 folds. The parameters of FS methods are tuned using Grid Search with a search range obtained empirically. Experiments were conducted using an NVIDIA RTX 3080 GPU, an Intel Core i7-12700 CPU, and 32 GB of RAM.

\subsection{Performance Analysis}
Fig. \ref{fig:box} illustrates the box plot of post-FS detection performance in terms of four different performance metrics. The distribution of recorded performance across different iterations of the cross-validation is shown by solid circles beside each box. In addition, the average and median of this distribution are indicated with a solid square and a red line, respectively. Furthermore, plus signs denote outliers in the plotted distribution. Maximum and minimum values are specified using green lines.

To study the post-FS detection performance, we select 110 features from the original pool of 3000 features for each method. The detection task is carried out using three classifiers, namely decision tree, random forest, and k-nearest neighbors. Throughout the experiments, the recorded performance metrics obtained by these classifiers are averaged to evaluate the performance of each FS algorithm.

Looking into the obtained accuracies in Fig. \ref{fig:box}(a), it can be implied that coupling classifiers with FS results in outstanding APT detection performance. Among the selected FS methods, CFMI generally outperforms the rest regardless of the performance metric. RFE, CAE, MI, GDS, SFS, and RliefF are ranked from second to seventh in terms of accuracy. Nevertheless, given the high imbalance ratio in the utilized dataset (see Table \ref{tab:attacks}), accuracy may accurately reflect the detection capability of this approach. Therefore, the F1-score of these methods is reported in Fig. \ref{fig:box}(b) to better understand the potency of FS when dealing with highly imbalanced data. While the obtained F1-scores are desirable, the ranking of FS algorithms has changed. Here, CAE outperforms RFE, and ReliefF results in a better F1-score than GDS.

In order to take false positives and false alarms into account, precision and recall associated with APT detection are additionally reported in Fig. \ref{fig:box}(c) and (d), respectively. Similar to the previous analysis, the obtained results in Fig. \ref{fig:box}(c) and (d) show superior performance for all FS techniques. The comparison of FS methods in terms of precision is very similar to that of F1-score. On the other hand, the ranking of methods when taking recall into consideration, is roughly the same as the comparison performed on the detection accuracy. This implies that the reason for the different ranking of algorithms when considering accuracy and F1-score is the degree of false alarm and false positive each of them leads to.

\subsection{Feature Analysis}

\begin{figure}
    \centering
    \rotatebox[origin=c]{270}{\includegraphics[trim={3.5cm, 0cm, 3.2cm, 0cm},clip,width=1.5\textwidth]{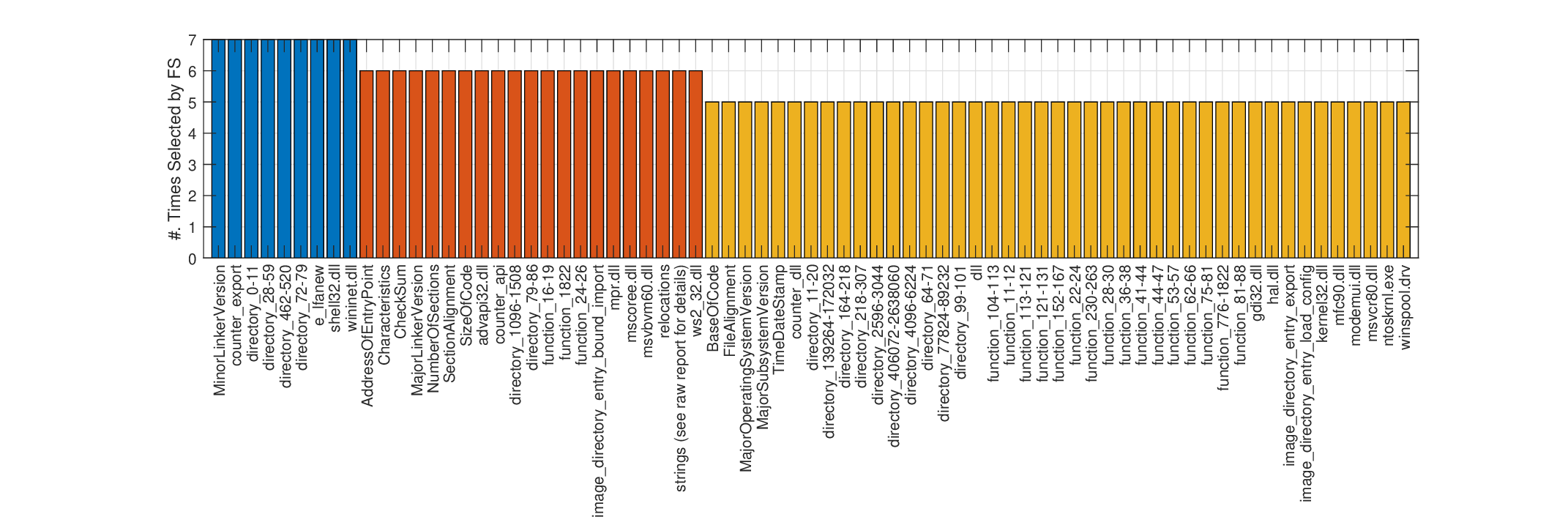}}
    \caption{The number of times different FS algorithms select each feature. Due to the abundance of features, only the top 73 features are reported. The rest of the features are selected by four or fewer FS algorithms.}
    \label{fig:bar}
\end{figure}

Fig. \ref{fig:bar} depicts the top 73 features selected by the majority of FS algorithms selected in this work. These features fall under different categories that relate to various aspects of binary's behavior and structure. Studying these categories helps with understanding how they contribute to detecting APTs within the network. Here, we categorize these features into three groups based on the level of priority.

The top nine features selected by all FS methods form the core of the detection strategy. These include winsock.dll, shell32.dll, and advapi32.dll Dynamic Link Library (DLL) imports and structural attributes such as MinorLinkerVersion and AddressOfEntryPoint. Because the features are present in all the selection procedures, it can be inferred that they are important aspects of APT?s behavior. As an example, wininet.dll and shell32.dll are frequently used by malicious programs for network communication and shell interaction. Advapi32.dll is used for performing registry and security operations which makes it a primary target for hackers trying to bypass and manipulate security protocols. AddressOfEntryPoint and MinorLinkerVersion are associated with the compilation and loading process of binaries, where abnormal alterations of the features may indicate the presence of malicious activities. These features suggest that both functional aspects and underlying structure indicators are beneficial in detecting malicious traffic associated with an APT.

The subsequent group of 21 features comprises, among others, additional DLLs and some functions that are substantially involved in the executable behavior. These include msvbvm60.dll, mscoree.dll, and ws2\_32.dll. Although these features are rather less important than the top group, they still provide opportunities to ascertain an APT presence. In addition, such features facilitate identifying the areas, resources, or libraries, that the binary code might communicate with. These features are also useful for operations that might be conducted at the system level. For instance, msvbvm60.dll is particular to the visual basic programs, and therefore its presence could indicate the usage of outdated programs that are security risks. mscoree.dll is used for .NET programs, and it is important to track since viruses may use the .NET runtime to facilitate their operations. The presence of ws2\_32.dll indicates possible networking operations, which can further be investigated for C2 communication or data exfiltration. Although not as central as the first nine features, this group is critical in narrowing down the broader depiction of what an executable does, in particular, the network-based-executables and managed code execution.

The rest of the features in Fig. \ref{fig:bar} are selected by five methods. These features are helpful in improving the detection system. Some of these are modemui.dll, hal.dll, ntdll.dll, as well as type and structure imports and function imports. These features seem to represent behavioral patterns that are more specific to a context or target, and may not be present in all APT attacks, but are still useful flags in certain scenarios. For example, modemui.dll and hal.dll belong to the class of hardware control and system management, vulnerabilities where sophisticated malware that aims to exploit even lower-level parts of a computer system may attack. Usage of ntdll.dll also aids in communication with the kernel, and unorthodox usage of this library may point towards alterations made at the executive level of the system catalog. Such features assist in exposing more of the subtle tactics and approaches that may not be noticeable from the previously discussed groups of features. Table \ref{tab:features} lists features that are selected by four or fewer FS algorithms. These features have lower priority compared to the features included in Fig. \ref{fig:bar} as they contribute less in detecting APT samples.

\begin{table}
\newcolumntype{C}{>{\centering\arraybackslash}X}
\newcolumntype{L}{>{\arraybackslash}p{10cm}}
\caption{List of selected features from the original APT dataset \cite{10.1007/978-3-319-60080-2_21} that contribute less compared to the top features mentioned in Fig. \ref{fig:bar}. Selected features are grouped based on the number of times they are picked by different FS methods, as inidcated in the left column.}
\begin{tabularx}{\textwidth}{CL}
\toprule
\#. Selected & Features\\
\midrule
\multirow{11}{*}{2} & netapi32.dll,
 image\_directory\_entry\_tls,
 directory\_12736-17864,
 ole32.dll,
 wtsapi32.dll,
 cabinet.dll,
 winmm.dll,
 e\_oeminfo,
 urlmon.dll,
 comdlg32.dll,
 directory\_53032-58000,
 directory\_2638060,
 NumberOfRvaAndSizes,
 azroles.dll,
 directory\_59-62,
 image\_directory\_entry\_delay\_import,
 directory\_11296-12736,
 directory\_172032-204800,
 imagehlp.dll,
 clbcatq.dll,
 directory\_24-28,
 e\_oemid,
 image\_directory\_entry\_security,
 setupapi.dll,
 resutils.dll,
 directory\_8396-11296,
 directory\_89232-114400,
 eappcfg.dll,
 directory\_420-462,
 opengl32.dll,
 oledlg.dll,
 imm32.dll,
 Machine,
 directory\_32632-36864,
 iphlpapi.dll,
 directory\_307-420,
 crypt32.dll,
 ImageBase,
 dciman32.dll,
 dnsapi.dll,
 oleaut32.dll.
 \\
 \midrule
     \multirow{16}{*}{3} & winsta.dll,
 rassapi.dll,
 e\_cs,
 directory\_204800-256116,
 PointerToSymbolTable,
 e\_magic,
 userenv.dll,
 directory\_256116-406072,
 uxtheme.dll,
 directory\_26208-32632,
 wintrust.dll,
 SizeOfHeapCommit,
 Subsystem,
 directory\_65536-77824,
 msvfw32.dll,
 other\_directory,
 e\_crlc,
 rassapi.dll,
 directory\_20-24,
 SizeOfStackCommit,
 e\_sp,
 function\_30-32,
 function\_38-41,
 function\_184-204,
 function\_263-303,
 function\_57-62,
 directory\_101-135,
 odbc32.dll,
 SizeOfImage,
 directory\_6224-8396,
 image\_directory\_entry\_debug,
 shlwapi.dll,
 MinorSubsystemVersion,
 msvcr90.dll,
 function\_141-152,
 function\_167-184,
 function\_88-96,
 function\_47-50,
 function\_542-776,
 function\_9-10,
 image\_directory\_entry\_exception,
 function\_303-357,
 function\_8-9,
 directory\_576-744,
 SizeOfInitializedData,
 cfgmgr32.dll,
 function\_50-53,
 directory\_135-164,
 msvcp90.dll,
 function\_32-34,
 function\_70-75,
 comctl32.dll,
 function\_96-104,
 dbnmpntw.dll,
 BaseOfData,
 cmdial32.dll,
 function\_424-542,
 function\_66-70,
 function\_204-230,
 image\_directory\_entry\_import,
 function\_131-141,
 cryptdll.dll.
 \\
 \midrule
     \multirow{14}{*}{4} & directory\_1508-2596,
 filesize,
 msvcrt.dll,
 image\_directory\_entry\_com\_descriptor,
 function\_12-15,
 oleacc.dll,
 function\_0-5,
 function\_19-22,
 function\_26-28,
 function\_357-424,
 image\_directory\_entry\_basereloc,
 Magic,
 DllCharacteristics,
 function\_15-16,
 image\_directory\_entry\_iat,
 directory\_988-1096,
 user32.dll,
 image\_directory\_entry\_resource,
 directory\_520-576,
 function\_5-6,
 function\_6-8,
 function\_10-11,
 functions,
 wsock32.dll,
 MinorOperatingSystemVersion,
 SizeOfHeaders,
 directory\_58000-65536,
 avicap32.dll,
 wnhelp.dll,
 secur32.dll,
 lz32.dll,
 directory\_744-988,
 msimg32.dll,
 mswsock.dll,
 mfc42.dll,
 ntdll.dll,
 psapi.dll,
 e\_cblp,
 directory\_71-72,
 winhttp.dll,
 directory\_86-99,
 SizeOfOptionalHeader,
 clusapi.dll,
 dsound.dll,
 esent.dll,
 SizeOfUninitializedData,
 MajorImageVersion,
 LoaderFlags,
 e\_ovno,
 version.dll,
 shimeng.dll,
 directory\_3044-4096,
 directory\_17864-21608,
 directory\_44944-53032.
 \\ 
\bottomrule
\end{tabularx}
\label{tab:features}
\end{table}

\section{Conclusion}
\label{sec:conclusion}
This paper designed a feature engineering framework to study the importance of features in the detection of APT samples based on the available public records of APTs. To ensure the generalization of our findings, 15 different APT cases were considered in our simulations. These samples were studied using several FS algorithms in combination with multiple classifiers. In addition, experiments were cross-validated using 10-fold to ensure statistical reliability of results. The feature analysis is performed by taking the results of all FS methods into consideration. The results show a superior APT detection performance when FS is incorporated into the detection process. After careful analysis of features, those identified among the top features were categorized into three groups based on the degree of importance. Finally, the necessary features for APT detection were explained along with the logic behind it. These findings provide insight for security experts into how data-driven APT detection can be enhanced using FS.


%
%
%
\bibliographystyle{splncs04}
\bibliography{main}

\begin{thebibliography}{10}
\providecommand{\url}[1]{\texttt{#1}}
\providecommand{\urlprefix}{URL }
\providecommand{\doi}[1]{https://doi.org/#1}

\bibitem{10.1007/978-3-030-61638-0_16}
Adepu, S., Palleti, V.R., Mishra, G., Mathur, A.: Investigation of cyber
  attacks on a water distribution system. In: Applied Cryptography and Network
  Security Workshops. pp. 274--291. Springer International Publishing, Cham
  (2020)

\bibitem{9947295}
Akbar, K.A., Wang, Y., Ayoade, G., Gao, Y., Singhal, A., Khan, L.,
  Thuraisingham, B., Jee, K.: Advanced persistent threat detection using data
  provenance and metric learning. IEEE Transactions on Dependable and Secure
  Computing  \textbf{20}(5),  3957--3969 (2023)

\bibitem{8606252}
Alshamrani, A., Myneni, S., Chowdhary, A., Huang, D.: A survey on advanced
  persistent threats: Techniques, solutions, challenges, and research
  opportunities. IEEE Communications Surveys \& Tutorials  \textbf{21}(2),
  1851--1877 (2019)

\bibitem{pmlr-v97-balin19a}
Bal{\i}n, M.F., Abid, A., Zou, J.: Concrete autoencoders: Differentiable
  feature selection and reconstruction. In: Proceedings of the 36th
  International Conference on Machine Learning. vol.~97, pp. 444--453 (Jun
  2019)

\bibitem{case2016analysis}
Case, D.U.: Analysis of the cyber attack on the ukrainian power grid.
  Electricity Information Sharing and Analysis Center (E-ISAC)
  \textbf{388}(1-29), ~3 (2016)

\bibitem{6198335}
Cheminod, M., Durante, L., Valenzano, A.: Review of security issues in
  industrial networks. IEEE Transactions on Industrial Informatics
  \textbf{9}(1),  277--293 (2013)

\bibitem{pmlr-v202-covert23a}
Covert, I.C., Qiu, W., Lu, M., Kim, N.Y., White, N.J., Lee, S.I.: Learning to
  maximize mutual information for dynamic feature selection. In: Proceedings of
  the 40th International Conference on Machine Learning. vol.~202, pp.
  6424--6447 (2023)

\bibitem{9563211}
Farajzadeh-Zanjani, M., Hallaji, E., Razavi-Far, R., Saif, M.:
  Generative-adversarial class-imbalance learning for classifying cyber-attacks
  and faults - a cyber-physical power system. IEEE Trans. on Dependable and
  Secure Computing  \textbf{19}(6),  4068--4081 (2022)

\bibitem{rfe}
Guyon, I., Weston, J., Barnhill, S., Vapnik, V.: Gene selection for cancer
  classification using support vector machines. Machine Learning
  \textbf{46}(1),  389--422 (2002)

\bibitem{9609642}
Hallaji, E., Farajzadeh-Zanjani, M., Razavi-Far, R., Palade, V., Saif, M.:
  Constrained generative adversarial learning for dimensionality reduction.
  IEEE Transactions on Knowledge and Data Engineering  \textbf{35}(3),
  2394--2405 (2023)

\bibitem{9207066}
Hallaji, E., Razavi-Far, R., Saif, M.: Detection of malicious scada
  communications via multi-subspace feature selection. In: International Joint
  Conference on Neural Networks (IJCNN). pp.~1--8 (2020).
  \doi{10.1109/IJCNN48605.2020.9207066}

\bibitem{HALLAJI2024103730}
Hallaji, E., Razavi-Far, R., Saif, M.: Expanding analytical capabilities in
  intrusion detection through ensemble-based multi-label classification.
  Computers \& Security  \textbf{139},  103730 (2024)

\bibitem{Hallaji2024}
Hallaji, E., Razavi-Far, R., Saif, M.: Robust federated learning for mitigating
  advanced persistent threats in cyber-physical systems. Applied Sciences
  \textbf{14}(19) (2024). \doi{10.3390/app14198840}

\bibitem{HASSANI2021104150}
Hassani, H., Hallaji, E., Razavi-Far, R., Saif, M.: Unsupervised concrete
  feature selection based on mutual information for diagnosing faults and
  cyber-attacks in power systems. Engineering Applications of Artificial
  Intelligence  \textbf{100},  104150 (2021)

\bibitem{HUANG2020101660}
Huang, L., Zhu, Q.: A dynamic games approach to proactive defense strategies
  against advanced persistent threats in cyber-physical systems. Computers \&
  Security  \textbf{89},  101660 (2020)

\bibitem{10.1007/978-3-319-60080-2_21}
Laurenza, G., Aniello, L., Lazzeretti, R., Baldoni, R.: Malware triage based on
  static features and public apt reports. In: Cyber Security Cryptography and
  Machine Learning. pp. 288--305. Springer International Publishing, Cham
  (2017)

\bibitem{10.1145/3386581}
Laurenza, G., Lazzeretti, R., Mazzotti, L.: Malware triage for early
  identification of advanced persistent threat activities. Digital Threats
  \textbf{1}(3), ~16 (aug 2020)

\bibitem{LI20218176}
Li, Y., Liu, Q.: A comprehensive review study of cyber-attacks and cyber
  security; emerging trends and recent developments. Energy Reports
  \textbf{7},  8176--8186 (2021)

\bibitem{1453511}
Peng, H., Long, F., Ding, C.: Feature selection based on mutual information
  criteria of max-dependency, max-relevance, and min-redundancy. IEEE
  Transactions on Pattern Analysis and Machine Intelligence  \textbf{27}(8),
  1226--1238 (2005)

\bibitem{9695986}
Rahman, Z., Yi, X., Khalil, I.: Blockchain-based ai-enabled industry 4.0 cps
  protection against advanced persistent threat. IEEE Internet of Things
  Journal  \textbf{10}(8),  6769--6778 (2023)

\bibitem{relief}
Robnik-{\v S}ikonja, M., Kononenko, I.: Theoretical and empirical analysis of
  relieff and rrelieff. Machine Learning  \textbf{53}(1),  23--69 (2003)

\bibitem{9907008}
Tran, N., Chen, H., Bhuyan, J., Ding, J.: Data curation and quality evaluation
  for machine learning-based cyber intrusion detection. IEEE Access
  \textbf{10},  121900--121923 (2022)

\bibitem{8417919}
Yang, L.X., Li, P., Yang, X., Tang, Y.Y.: A risk management approach to
  defending against the advanced persistent threat. IEEE Transactions on
  Dependable and Secure Computing  \textbf{17}(6),  1163--1172 (2020)

\bibitem{8941015}
Yang, L.X., Li, P., Yang, X., Xiang, Y., Jiang, F., Zhou, W.: Effective
  quarantine and recovery scheme against advanced persistent threat. IEEE
  Transactions on Systems, Man, and Cybernetics: Systems  \textbf{51}(10),
  5977--5991 (2021)

\bibitem{10163904}
Yu, Z., Gao, H., Cong, X., Wu, N., Song, H.H.: A survey on cyber–physical
  systems security. IEEE Internet of Things Journal  \textbf{10}(24),
  21670--21686 (2023)

\bibitem{10.1145/1273496.1273641}
Zhao, Z., Liu, H.: Spectral feature selection for supervised and unsupervised
  learning. In: Proceedings of the 24th International Conference on Machine
  Learning. p. 1151–1157 (2007)

\bibitem{9923774}
Zhu, T., Ye, D., Cheng, Z., Zhou, W., Yu, P.S.: Learning games for defending
  advanced persistent threats in cyber systems. IEEE Transactions on Systems,
  Man, and Cybernetics: Systems  \textbf{53}(4),  2410--2422 (2023)

\end{thebibliography}

\end{document}